\documentclass{llncs}

\usepackage{graphicx}
\usepackage{tabularx}
\begin{document}

\newcolumntype{Z}{>{\centering\arraybackslash}X} 

\mainmatter              

\title{Dynamic Top-$k$ Dominating Queries}
\titlerunning{Dynamic Top-$k$ Dominating Queries}  
%
\author{Andreas Kosmatopoulos \and Kostas Tsichlas}
%
\authorrunning{Kosmatopoulos and Tsichlas} 
%
\tocauthor{Andreas Kosmatopoulos and Kostas Tsichlas}
\institute{Informatics Department, Aristotle University of Thessaloniki, Greece\\
\email{\{akosmato, tsichlas\}@csd.auth.gr}}

\maketitle              

\begin{abstract}
Let $\mathcal{S}$ be a dataset of $n$ $2$-dimensional points. The top-$k$ dominating query aims to report the $k$ points that dominate the most points in $\mathcal{S}$. A point $p$ dominates a point $q$ iff all coordinates of $p$ are smaller than or equal to those of $q$ and at least one of them is strictly smaller. The top-$k$ dominating query combines the dominance concept of maxima queries with the ranking function of top-$k$ queries and can be used as an important tool in multi-criteria decision making systems. In this work, we propose novel algorithms for answering semi-dynamic (insertions only) and fully dynamic (insertions and deletions) top-$k$ dominating queries. To the best of our knowledge, this is the first work towards handling (semi-)dynamic top-$k$ dominating queries that offers algorithms with asymptotic guarantees regarding their time and space~cost.
\end{abstract}

\section{Introduction} \label{section:Introduction}
In recent years, there has been an increasing research interest in preference queries, due to their ability to select the most interesting objects of a given dataset. The dataset objects are characterized by a number of often contradictory attributes; thus selecting a suitable subset becomes a challenging task.

\begin{figure}
\centering
\includegraphics[width=0.43\textwidth]{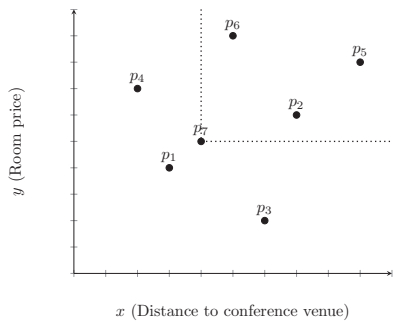}
\vspace{-0.4cm}
\caption{Hotel Dataset} \label{fig:Hotels}
\vspace{-0.6cm}
\end{figure}

As an example, consider a dataset containing hotels which will accommodate researchers during a conference. Assume that each hotel is represented as a $2$-dimensional point of two attributes: its distance from the conference venue and its room price. Generally, a potential customer would be interested in hotels that have both of these attributes as minimized as possible. The solution would then consist of all the hotels that are in a sense more ``preferred'' than the others. There have been three different kinds of preference queries analyzed in literature which we discuss below. For the remainder of this introductory section we assume that $\mathcal{S}$ is a dataset of points in the $2$-dimensional plane $\mathcal{R}^2$. Furthermore, Figure~\ref{fig:Hotels} illustrates a dataset of hotels that will serve as a running example.

A \emph{Top-$k$ query}~\cite{FaginLN01} provides a preference ranking for all the objects in $\mathcal{S}$ using a user-defined ranking function. Let $\mathcal{F}:\mathcal{R}^2\rightarrow\mathcal{R}$ be a monotone ranking function used to assign a value to each point in $\mathcal{S}$. The top-$k$ query will return the $k$ points in $\mathcal{S}$ with the smallest $\mathcal{F}$ value. For example, in Figure~\ref{fig:Hotels} the top-$2$ points for $\mathcal{F}=x+y$ are $p_1$ and $p_3$. We assume that $k$ is small when compared to the dataset size (imagine a researcher asking for the top-$400$ among the $535$ hotels in New York!). An advantage of top-$k$ queries is that the user-defined parameter $k$ controls the output size. However, they are based on a preference function and as such different user-defined ranking functions may produce different rankings.

A \emph{Maxima (Skyline) query}~\cite{BorzsonyiKS01} returns all the points in the dataset that are not dominated by any other point. A point $p\in\mathcal{S}$ dominates a point $p'\in\mathcal{S}$ if all the coordinates of $p$ are smaller than the coordinates of $q$ and at least one of the coordinates of $p$ is strictly smaller than the respective coordinate of $q$. The above definition assumes that smaller values are preferable to larger at all dimensions. For datasets where larger values are preferable, the definition may be altered accordingly. As an example, the maximal points of Figure~\ref{fig:Hotels} are $p_1$, $p_3$ and $p_4$. The major advantage of maxima queries is that they are based on the dominance concept and as a result, they do not require a user-defined preference ranking function. Furthermore, due to the definition of the maxima query, the results are not affected by potentially different scales at different dimensions. On the other hand, maxima queries do not control the size of their output and in extreme cases it can be as large as the dataset size.

A \emph{Top-$k$ Dominating query}~\cite{Papadias05} combines the dominance concept of maxima queries with the ranking function of top-$k$ queries. More specifically, each point $p\in\mathcal{S}$ is assigned a dominance score which is equal to the number of points it dominates. A top-$k$ dominating query returns the $k$ points in $\mathcal{S}$ with the highest dominance score. For instance, the top-$2$ dominating points of Figure \ref{fig:Hotels} are $p_1$ ($score = 4$) and $p_7$ ($score = 3$). Top-$k$ dominating queries provide an intuitive way of ranking a dataset's objects by using their dominance relationship instead of a user-defined ranking function. In addition, their output size is bounded by the parameter $k$ and they are unaffected by different data scales at different dimensions. In a sense, they combine the merits of maxima and top-$k$ queries. Consequently, top-$k$ dominating queries can be used as an important tool in multi-criteria optimization applications (e.g. in \cite{SkoutasSSKS09} they use a form of the dominance concept to rank web services under multiple criteria).

We define a query $q$ in $\mathcal{S}$ to be decomposable~\cite{BenSax80} if its output can be computed accurately by performing $q$ in a partition of $\mathcal{S}$. In contrast to top-$k$ and maxima queries, a top-$k$ dominating query is a \emph{non-decomposable query} since the score of each point is dependent on the coordinates of all the other points in $\mathcal{S}$. This fact greatly increases the difficulty of the problem.

In this work, we propose a novel solution for semi-dynamic and fully dynamic top-$k$ dominating queries on a dataset $\mathcal{S}$ of $2$-dimensional points, where $k$ is a user-defined parameter that is fixed between queries. In the semi-dynamic setting we support the insertion of new points, while in the fully dynamic setting we further support the deletion of existing points. To the extent of our knowledge, this is the first work that offers asymptotic time and space bounds for these particular queries. Note, that this paper concentrates on $2$-dimensional data for two reasons. First, there is no previous work with asymtpotic guarantees and as a result, this paper provides a deeper understanding of the complexity of the problem. The second, more practical, reason is that many applications are inherently $2$-dimensional. This is because, one often faces the situation of having to strike a balance between a pair of naturally contradicting
factors (e.g., price vs quality, space vs query time).

In the remainder of the introductory section we review related work and present our contributions. In Section~\ref{section:Preliminaries} we provide an overview of the basic concepts that will be used throughout the rest of this work. In Sections \ref{section:SemiDynTopKDom} and \ref{section:DynamicTopKDom} we formally describe our proposed solution for the semi-dynamic and fully dynamic top-$k$ dominating query, respectively. Finally, in Section \ref{section:ConclusionsFutureWork} we conclude and offer directions for future work.



\subsection{Related Work and Our Results} \label{subsection:RelatedWork}
The simplest approach to answering the top-$k$ dominating query would to be to compare each point $p$ with every other point $q$ in the dataset and increment $p$'s score if it dominates $q$. This results in $O(n^{2})$ time cost and $O(n)$ space cost. An approach with lower time complexity would be to use a $2$-dimensional range counting data structure. Assuming word size $w=\Omega{(\log n)}$, for each point $p:(x_p,y_p)$ in a dataset $\mathcal{S}$, the points lying in the query rectangle $Q=\left[ {{x_p},{\infty}} \right) \times \left[ {y_p,\infty } \right)$ can be counted in $O((\frac{\log n}{\log\log n})^2)$ time and $O(n)$ space using the method described in \cite{HeMunroWADS11}. The number of points found in $Q$ is equal to $p$'s dominance score. In order to compute the dominance score of each point, we repeat the process for all the points in $\mathcal{S}$ in $O(n(\frac{\log n}{\log\log n})^2)$ total time. Insertions and deletions can be trivially supported in $O(n)$ time since one has to update the dominance scores of all points in the worst-case. In the following, we describe more elaborate methods to answer a top-$k$ dominating query.

Papadias et al.~\cite{Papadias05}, first proposed the $d$-dimensional top-$k$ dominating query along with a solution based on the iterative computation of a dataset's maxima points. More specifically, they observed that the top-$1$ dominating point of a dataset is contained in the dataset maxima's points. This stems from the observation that for every point $p$ not in the maxima, there exists a point $p'$ in the maxima that dominates it and, as a result, $p'$ has a larger score than $p$. Thus, in their approach, they compute the set of maxima $M$ (using the BBS algorithm \cite{Papadias05}) and compute the dominance score of all the points in $M$. The point $q$ with the highest score is the top-$1$ dominating point and is thereby reported. Finally, $q$ is removed from the dataset and the procedure is repeated until $k$ points have been reported. Since in the worst case the set of maxima may be the whole dataset, the algorithm requires $O(n^2)$ time and $O(n)$ space\footnote{The BBS algorithm is based on the use of R-trees which require linear space}.

Yiu and Mamoulis~\cite{YiuM09} suggested using aggregate R-trees (aR-trees) to efficiently compute $d$-dimensional top-$k$ dominating queries. They provided various algorithms based on aR-trees that proved experimentally to be quite fast. They also make an analytic study making the assumption that the data points are uniformly and independently distributed in a domain space. The authors do not make any statement for the worst-case time complexity of the query but it is certainly $\Omega(n)$.

 Both methods \cite{Papadias05,YiuM09} focus on the top-$k$ dominating query, where $k$ is arbitrary. Update operations can be applied in both cases with a linear time cost. However, the top-$k$ dominating query has to be re-evaluated in both cases. Finally, both prove the efficiency of their approach experimentally (extensive experiments can be found in~\cite{YiuM09}).
\vspace{-0.5cm}
\begin{table}
\centering\footnotesize
\caption{SD stands for Semi-Dynamic, where only insertions are supported while FD stands for Fully-Dynamic where both insertions and deletions are supported. \vspace{-0.3cm}}\label{table:OurResults}
\begin{tabular}{l|c|c|c}
\hline
Algorithm & Preprocessing & Query & Update (amortized)\\
\hline
SD/$k$-list & $O(n(\frac{\log n}{\log\log n})^{2})$ & $O(k)$ & $O((\frac{\log n}{\log\log n})^{2}+k^{2}\log n)$ \\
SD/$1$-list & $O(n(\frac{\log n}{\log\log n})^{2})$ & $O(k\log n)$ & $O((\frac{\log n}{\log\log n})^{2}+k\log n)$ \\
FD/$k$-list & $O(n(\frac{\log n}{\log\log n})^{2})$ & $O(k)$ & $O(\sqrt{n}(\frac{\log n}{\log\log n})^{2}+(k+\sqrt{n})k\log n)$ \\
FD/$1$-list & $O(n(\frac{\log n}{\log\log n})^{2})$ & $O(k\log n)$ & $O(\sqrt{n}(\frac{\log n}{\log\log n})^{2}+(k+\sqrt{n})\log n)$ \\
\hline
\end{tabular}
\end{table}

\vspace{-0.5cm}
\noindent In this paper we tackle for the first time the problem of answering dynamic top-$k$ dominating queries providing a complexity analysis of all supported operations. Our algorithms work well under the assumption that $k$ is a fixed user specified parameter which is small when compared to the size $n$ of the dataset. We also attack the problem in two dimensions for reasons previously mentioned. Finally, our algorithms are based on a novel restricted dynamization of layers of maxima~\cite{BlunckVAlgorithmica10}. This is of independent interest in case we only need to access the first $k$ layers of maxima.

We consider the problem in the semi-dynamic case (insertions only), where logarithmic  complexities are attained. However, in the fully-dynamic case, we are only able to attain polynomial complexities for update operations. For each case we provide two solutions ($k$-list and $1$-list) that provide a trade-off between update and query time. All our algorithms use linear space. Table \ref{table:OurResults} provides a detailed  overview of our results.


\section{Preliminaries} \label{section:Preliminaries}

Let $\mathcal{S}$ be a set of $n$ $2$-dimensional points $p_i$ where $p_i=(x_{i},y_{i}),1\leq i\leq n$ with $x_{i}$ being the $x$-coordinate of $p_i$ and $y_{i}$ the $y$-coordinate of $p_i$. A point $p=(x_{p},y_{p})\in\mathcal{S}$ \emph{dominates} another point $q=(x_{q},y_{q})\in\mathcal{S}$ iff $(x_{p} \leq x_{q},y_{p}<y_{q})\vee(x_{p} < x_{q},y_{p}\leq y_{q})$. We use ($p \succ q$) to denote that $p$ dominates $q$. The \emph{dominance score} of a point $p$ is equal to $s(p) = |\{q \in \mathcal{S}| p \succ q\}|$. We augment the definition of each point $p_i$ to also include its score $s_i$.
Thus, a \emph{top-$k$ dominating query} in $\mathcal{S}$ aims to report the $k$ points of $\mathcal{S}$ with the highest dominance score.

The algorithms presented in the remaining sections are based on the concept of \emph{layers of maxima}. In order to compute the layers of maxima for a dataset $\mathcal{S}$ we perform a maxima query on $\mathcal{S}$, remove the answer set of points from $\mathcal{S}$ and repeat the process until no points remain in $\mathcal{S}$. The set that results from the $i$-th maxima query forms the $i$-th layer of maxima. By collecting all the layers, we form the layers of maxima of $\mathcal{S}$.

\section{Semi-Dynamic Top-$k$ Dominating Points} \label{section:SemiDynTopKDom}
In this section we propose a solution to the semi-dynamic top-$k$ dominating query problem and describe in detail the data structures and algorithms we use to achieve it.
Let $\mathcal{S}$ be a set of $n$ $2$-dimensional points. Recall that the semi-dynamic top-$k$ dominating query aims to report the $k$ points in $\mathcal{S}$ with the highest dominance score where $k$ is a fixed user-provided parameter. Furthermore, $\mathcal{S}$ is subject to insertions of new points. This poses an additional challenge since after inserting a new point, it is possible that the dominance score of many (or even all) the points in $\mathcal{S}$ must be updated. Individually updating the score of each such point would be computationally prohibitive so we follow a different approach and only update lazily the score of groups of points that are candidates for being in the final answer.

We first note that when a point $p$ dominates another point $q$, $p$'s score is strictly greater than the score of $q$:
\vspace{-0.2cm}
\begin{equation}\label{eq:DomScoreProperty}
\forall p,q \in \mathcal{S}, p\succ q \Rightarrow score_p > score_q
\end{equation}

\vspace{-0.2cm}
Organizing $\mathcal{S}$ into layers of maxima offers an intuitive way of using the above property to eliminate points that are not possible to belong in the final answer. As an example, consider a top-$1$ dominating query in $\mathcal{S}$. The point with the highest dominance score is found in the first layer of maxima of $\mathcal{S}$ since all the points in the second and subsequent layers are dominated by at least one other point. Similarly, in a top-$2$ dominating query, the first point is found in the first layer and the second point is found in either the first or the second layer. In general, the following lemma holds for the top-$k$ dominating points (see Appendix~\ref{app:lemma1} for the proof):

\begin{lemma}\label{lemma:FirstKLayers}
The top-$k$ dominating points of $\mathcal{S}$ are found in the first $k$ layers of maxima of $\mathcal{S}$.
\end{lemma}

A direct consequence of Lemma \ref{lemma:FirstKLayers} is that, when inserting a new point $p$, we only need to update the scores of some points in the first $k$ layers of maxima. However, some of the layers may have many points and thus individually updating the score of these points would result in a high update cost. To avoid this, after inserting a new point $p$ in $\mathcal{S}$, we find only the first and last point that dominate $p$ in each layer. This pair of points denotes an interval that marks all the points in each layer whose score must be updated. Consequently, by examining only $O(1)$ points in each layer the total update cost is reduced.

Lastly, an issue brought up by the use of layers of maxima is that the insertion of a new point $p$ may create cascading changes to the structure of the layers. In particular, by inserting $p$ into $\mathcal{S}$, $p$ must also be inserted in one of the layers of $\mathcal{S}$. Let $L_i$ be that layer. The insertion of $p$ in $L_i$ may cause some of its points to be discarded as a result of them being dominated by $p$. This group of points must be inserted into the next layer $L_{i+1}$ possibly discarding some of the points in $L_{i+1}$ in the process. Due to Lemma \ref{lemma:FirstKLayers} and the fact that only insertions are allowed, this chain of operations only has to be performed up until the $k$-th layer.

To achieve efficient insertion, we model each layer as an $(a,b)$-tree. In the following, we provide a detailed overview of the data structure and the operations it supports and then we describe the update and query algorithms for the semi-dynamic top-$k$ dominating query.

\subsection{The augmented $(a,b)$-tree} \label{subsection:AugABTree}
We use an augmented leaf-oriented $(a,b)$-tree to model each layer of maxima. Assume that $L$ is a layer of maxima containing $m$ points, i.e. $p_1, p_2,\ldots p_m$ where $p_i=(x_{i},y_{i},s_i),1\leq i\leq m$. Since the points in $L$ are totally ordered on each dimension\footnote{For two points $p_a=(x_{a},y_{a},s_a)$ and $p_b=(x_{b},y_{b},s_b)$ iff $x_{a} > x_{b}$ then $y_{b} > y_{a}$}, we can use a single $(a,b)$-tree to search among the points in both dimensions. To achieve that, each inner node stores representative keys for both dimensions, instead of storing keys for only one of them.

For each node $v$ of the tree with $height \geq \log_{b}{k}$ we maintain a field $add(v)$. The field's contents denote a score that has to be added to the score of all the points in $v$'s subtree. Finally, each node $v$ with $height \geq \log_{b}{k}$ is augmented with an $k$-sized list $top(v)$ which stores the $k$ points with the highest score in $v$'s subtree\footnote{The height is measured from the leaves to the root of the tree}. The points in $top(v)$ are sorted according to their score. 
For the remainder of this work we assume that $b=O(1)$ since we present main memory algorithms. The following lemma provides the tree's total space cost (for a proof see Appendix~\ref{app:lemma2}).
\begin{lemma} \label{lemma:AugABTreeTotalSpace}
The total space required by an augmented $(a,b)$-tree storing $m$ points is $O(m)$.
\end{lemma}

The following lemma provides the time complexity for the construction of an augmented $(a,b)$-tree over $m$ points (for a proof see Appendix~\ref{app:construction}).
\begin{lemma} \label{lemma:construction}
The construction of an augmented $(a,b)$-tree over $m$ points can be carried out in $O(m\log{k})$ time, where $k$ is a user-defined parameter.
\end{lemma}

\vspace{-0.5cm}
\subsubsection{Operations.} \label{subsubsection:AugABTreeOperations}
In this section all the operations supported by the augmented $(a,b)$-tree are formally described. More specifically, the augmented $(a,b)$-tree supports searching for a point, inserting a new point, or deleting an existing one. Furthermore, splits and concatenations between two different $(a,b)$-trees are also supported.

The search operation \texttt{search$(T,p_z)$} locates in the augmented $(a,b)$-tree $T$ a specific point $p_z$ and can be performed with respect to either dimension of $p_z$ by using the appropriate set of keys. Let $v$ be a node of $T$, $x_1, x_2, \ldots, x_{b-1}$ be the $x$-representative keys of $v$'s children and $y_1, y_2, \ldots, y_{b-1}$ be the $y$-representative keys of $v$'s children. In order to search for a point $p_z=(x_{z},y_{z},s_z)$ in $T$, we begin at the root and search down until we reach a leaf. If the search is performed on the $x$ dimension, we select the $i$-th child of $v$ such that $x_{i-1} < x_z \leq x_i$. Otherwise, if the search is performed on the $y$ dimension, we select the $i$-th child of $v$ such that $y_{i-1} > y_z \geq y_i$. Since $T$ is height-balanced, a search operation requires $O(b\log m)=O(\log m)$ time.

The rest of the operations are based on node splits and node merges. For reasons of clarity, we first describe how node splits and node merges are handled on the augmented $(a,b)$-tree in relation to typical $(a,b)$-trees.

The node split operation \texttt{node\_split$(v,v_1,v_2)$} is performed similarly to the split operation of typical $(a,b)$-trees with a few modifications. More specifically, before dividing a node $v$ into two nodes $v_1$ and $v_2$ we check the contents of $add(v)$. If $add(v)$ stores a value different than $0$, we add the contents of $add(v)$ to the $add$ variable of $v$'s children and set $add(v)$ to $0$. Afterwards, $v$ is divided into $v_1$ and $v_2$ and the keys for the $x$ and $y$ dimensions of $v$ are ``shared'' between $v_1$ and $v_2$ in $O(b)=O(1)$ time. After sharing the keys, $top(v_1)$ for $v_1$ and $top(v_2)$ for $v_2$ must be recomputed. If $height_v > \log_{b}k$ we can compute $top(v_1)$ and $top(v_2)$ in $O(k)$ time by simultaneously traversing the $O(b)=O(1)$ $top$ lists of $v_1$'s and $v_2$'s children respectively. As a result, the split operation requires $O(k)$ time in this case.

If $height_v = \log_{b}{k}$ then the children of $v_1$ and $v_2$ are not augmented with $top$ lists and thus computing $top(v_1)$ and $top(v_2)$ cannot be performed using the above procedure. In this case, we can compute $top(v_1)$ and $top(v_2)$ by simply traversing the $top(v)$ list and assigning each of its points $p_i$ to either $top(v_1)$ or $top(v_2)$. In order to do this, we first find the child $v_c$ of $v$ that contains $p_i$ in $O(b)=O(1)$ time using $v$'s representative keys. Afterwards, we discover whether $v_c$ is a child of $v_1$ or $v_2$ using the representative keys of $v_1$ and $v_2$, and then assign $p_i$ to $top(v_1)$ or $top(v_2)$ respectively. Since these steps are repeated for each point in $top(v)$ the split operation totally requires $O(k)$ time. Finally, if $height_v < \log_{b}{k}$ then the split operation requires $O(b)=O(1)$ time since $v$ is not augmented with $top(v)$. In conclusion, a split operation is performed in $O(k)$ time in every case.



For the merge operation \texttt{node\_merge$(v_1,v_2,v)$} we follow a similar procedure to the merge operation of standard $(a,b)$-trees. More specifically, before merging two nodes $v_1$ and $v_2$ into $v$ we check the contents of $add(v_1)$ and $add(v_2)$. If $add(v_1)$ stores a value different than $0$ we add the contents of $add(v_1)$ to $v_1$'s children and set $add(v_1)$ to $0$. We follow the same procedure for $add(v_2)$. Then, $v_1$ and $v_2$ are merged into $v$ and the keys for the $x$ and $y$ dimensions of $v$ are derived from the keys of $v_1$ and $v_2$ in $O(b)=O(1)$ time. After merging the keys, $top(v)$ must be recomputed. To achieve this, we simultaneously traverse $top(v_1)$ and $top(v_2)$ and store the $k$ points with the highest score in $top(v)$ in $O(k)$ time.
As in the split operation, if $height_v < \log_{b}{k}$ the merge operation requires $O(b)=O(1)$ time due to the fact that $v$ is not augmented with $top(v)$. As a result, the merge operation requires $O(k)$ time in all cases.

Operation \texttt{insert$(T,p)$}, inserts a point $p$ in $T$. The point is inserted as a leaf in $T$ and the tree is rebalanced using node splits. Since there are $O(\log{m})$ node splits the time cost to insert a point is $O(k\log{m})$.

Operation \texttt{delete$(T,p)$}, removes a point $p$ from $T$. The leaf corresponding to the point is removed and the resulting tree is rebalanced. There are $O(\log m)$ merges and a possible terminating split and as a result the time cost to delete a point is $O(k\log{m})$.

Using the node split and node merge operations as building blocks, we can define two additional operations on the augmented $(a,b)$-trees: Tree Concatenation and Tree Split. For both the operations, we use the definition and algorithms provided in \cite{Mehlhorn84}.

Operation \texttt{ConCat$(T_1,T_2,T_3)$}, concatenates two augmented $(a,b)$-trees $T_1$ and $T_2$ into a third augmented $(a,b)$-tree $T_3$. This operation can only be performed if $\max\{T_1\} \leq~ \min\{T_2\}$. In a tree concatenation there is one merge operation and up to $O(\log{\max(|T_1|,|T_2|)})$ split operations performed. Since merge and split operations cost $O(k)$ time, a tree concatenation operation requires $O(k\log{\max(|T_1|,|T_2|)})$ time.

Operation \texttt{Split$(T_1,val,T_2,T_3)$}, splits an augmented $(a,b)$-tree $T_1$ into two augmented $(a,b)$-trees $T_2$ and $T_3$ at element $val$ with respect to the one of the two dimensions, so that $T_2\leftarrow\{ z\in T_1; z\leq val\}$ and $T_3\leftarrow\{ z\in T_1; z> val\}$. In a tree split operation the starting $(a,b)$-tree is first split into two forests of trees. Then, the roots of the trees in each forest are merged with each other recursively. Splitting the tree into two forests requires $O(\log |T_1|)$ time and since there are $O(\log |T_1|)$ merges for both forests, each requiring $O(k)$ time, a tree split operation requires $O(k\log |T_1|)$ time. The following theorem summarizes the discussion on the $(a,b)$-tree.

\begin{theorem} \label{theorem:AugABTreeCombined}
Given $m$ $2$-dimensional points $p_i = (x_i,y_i,s_i)$ where $1 \leq i \leq m$ and a parameter $k$, we can construct in $O(m\log{k})$ time an augmented $(a,b)$-tree $T_1$ that uses $O(m)$ space, and supports \texttt{search$(T_1,p)$} in $O(\log{m})$ time, \texttt{insert$(T,p)$} and \texttt{delete$(T,p)$} in $O(k\log{m})$ time and \texttt{Split$(T_1,val,T_2,T_3)$} in $O(k\log{m})$ time. Furthermore, given an augmented $(a,b)$-tree $T_4$ where $\max\{T_1\} \leq \min\{T_4\}$, \texttt{ConCat$(T_1,T_4,T_5)$} is supported in $O(k\log{\max(|T_1|,|T_4|)})$ time.
\end{theorem}

\subsection{Insertion} \label{subsection:Insertion}
Let $p=(x_p,y_p,s_p)\in\mathrm{R}$ be a point to be inserted into $\mathcal{S}$. Furthermore, let $L_1, \ldots, L_k$ be the first $k$ layers of maxima of $\mathcal{S}$. Before inserting $p$ we compute its dominance score using the dynamic range counting data structure proposed in \cite{HeMunroWADS11}\footnote{The data structure is built only once as a preprocessing step before the first insertion}. The data structure supports queries in $O((\frac{\log n}{\log\log n})^{2})$ worst-case time and insertions and deletions in $O((\frac{\log n}{\log\log n})^{2})$ amortized time (assuming word size $w=\Omega(\log n)$ in the word RAM model).

Afterwards, we find if $p$ must be inserted in one of $L_1, \ldots, L_k$ by searching each of the $k$ respective $(a,b)$-trees for $p$. Starting from $L_1$ and iterating towards $L_k$, we search each tree for $p$ both in the $x$ and in the $y$ dimension and retrieve the predecessor of $p$ in the $x$ dimension and the successor of $p$ in the $y$ dimension. If neither of those two points dominate $p$, we insert $p$ in the tree's respective layer and stop the iteration. Otherwise, the iteration may end without any layer satisfying the above condition. In that case, $p$ does not become a member of any of the $k$ first layers.

If we do not insert $p$ in any of the $k$ first layers then we only have to update the scores of some points in each of $L_1, \ldots, L_k$. Otherwise, assume that $p$ is inserted into $L_i$ where $1 \leq i \leq k$. Then we have to update scores of points in $L_1, \ldots, L_{i-1}$ and alter the structure of $L_i, \ldots, L_k$. We first describe how to handle \emph{score updating} on a layer and afterwards how to \emph{alter a layer's structure} using tree splits and tree concatenations.


To update the score of the points in a layer $L$, we perform this procedure. We search the augmented $(a,b)$-tree of $L$ for $x_p$ and $y_p$. All points whose score must be updated lie to the left of $x_p$ and to the right of $y_p$. Since updating the score of each point would be time consuming, we only find the two boundary points that define the above interval and mark the subtrees between them.

We start from $height = \log_{b}k + 1$ of the two search paths and move up towards the root, adding $+1$ to $add(v)$ if $v$ is a node hanging to the left of the search path for $x_p$ or to the right of the search path for $y_p$. Using this method we denote that the score of all the points in $v$'s subtree must be incremented by one, without actually visiting the points themselves. Adding $+1$ to $add(v)$ does not change $top(v)$ since we increment the score of all the points in $v$'s subtree and thus their relative order according to score remains unchanged. For each node $v'$ with $height_{v'} = \log_{b}k$, instead of incrementing $add(v')$, we exhaustively check the points in $list(v')$ and individually update their score based on if they are dominating $p$. For each node with $height < \log_{b}k$ no action is necessary since all the points in its subtree can be found in the $top$ list of its ancestor with $height = \log_{b}k$. Thus, at the end, we have indirectly marked all the points between $y_p$ and $x_p$ for score increment.

Finally, we update the $top$ lists of the nodes in the search path as a result of modifying the $add$ fields of their children. Starting from $height = \log_{b}k + 1$ and moving towards the root, we recursively compute the $top$ list of each node $v$ by simultaneously merging the $top$ lists of its children. While merging the lists, we also add the contents of each node's $add$ field to the score of the node's $top$ list points so as to take into account the score changes caused by the insertion of $p$. At the end, the $top$ list found in the root of the $(a,b)$-tree will have the correct top-$k$ points for that layer of maxima.

Updating the $add$ fields requires $O(\log n)$ total time while merging the top-$k$ lists of a node's children requires $O(bk)=O(k)$ time and as a result the total cost for all the nodes in the search paths of the tree is $O(bk\log n)=O(k\log n)$ time. Thus, the total time required to update scores in a layer is $O(k\log n)$.

On the second case, the point $p$ may have to be inserted in a layer of maxima $L_i$. Since inserting or deleting points from the layer of maxima one-by-one would be time consuming, we insert the point and remove the now-dominated points with a series of tree splits and tree concatenations. First, we find the interval as previously by querying the layer of maxima tree $T$ for $x_p$ and $y_p$. Then we perform the following sequence of operations in order: 1) \texttt{Split$(T,y_p,T_1,T_2)$}, 2) \texttt{Split$(T_2,x_p,T_3,T_{discard})$}, 3) \texttt{insert$(T_3,p)$} and 4) \texttt{ConCat$(T_3,T_1,T_{new})$}.
The layer of maxima tree $T_{new}$ for $L_i$ now correctly has $p$ inserted and every point previously in $L_i$ that is now dominated by $p$ (i.e. $T_{discard}$) has been discarded.

The tree $T_{discard}$ is now propagated to the next layer of maxima $L_{i+1}$ where we repeat the above procedure with $T_{discard}$ as the input. Since $T_{discard}$ may have more than one points, instead of inserting them one-by-one we perform a tree concatenation at step (3) instead of an insertion. Finally, the insertion spot of $T_{discard}$ in $L_{i+1}$ can be found by querying the tree of $L_{i+1}$ for $p'=(x_p',y_p')$ where $x_p'$ is the $x$ coordinate of the leftmost point in $T_{discard}$ and $y_p'$ is the $y$ coordinate of the rightmost point in $T_{discard}$. In each layer of maxima we perform a series of $O(1)$ splits and concatenations. Thus, the total time required to alter a layer's structure given a point or an $(a,b)$-tree as an input is $O(k\log n)$.

As described in the beginning of the section, after an insertion a layer must either update the score of some of its points or alter its structure. Since either case requires $O(k\log n)$ time, the time cost of manipulating the $k$ first layers after an insertion is $O(k^{2}\log n)$. Adding the cost of maintaining the dynamic range counting data structure, the total insertion cost is $O((\frac{\log n}{\log\log n})^{2}+k^{2}\log n)$ amortized time (assuming word size $w=\Omega(\log n)$).

\subsection{Query} \label{subsection:Query}
To find the top-$k$ dominating points of $\mathcal{S}$, we apply Lemma \ref{lemma:TopKDomLemma} on all the top-$k$ lists found in the root of each $(a,b)$-tree of each of the $k$ first layers of maxima. Let $I$ be the list returned by Lemma \ref{lemma:TopKDomLemma}. By selecting the $(|I|-k+1)$-th order statistic of $I$ we obtain the dominance score $\tau$ of the $k$-th top dominating point. Finally, we traverse all the top-$k$ lists we previously collected and report all points with score larger than $\tau$. Since the lists are sorted according to their score, we can stop traversing a list when a point with score lower than $\tau$ has been found. Applying Lemma \ref{lemma:TopKDomLemma} requires $O(k)$ time while finding the $(|I|-k+1)$-th order statistic of $I$ requires $O(I)=O(k)$ time. Finally, traversing all requires $O(I)=O(k)$ time in total. By combining all of the above, we achieve $O(k)$ query time.

\subsection{Reducing the Update Cost} \label{subsection:ReduceUpdateCost}
We can reduce the algorithm's update cost by shrinking the size of the $top$ list in each node of each $(a,b)$-tree. In particular, instead of storing $k$ points in each $top$ list we only store $1$. This removes the cost of computing $top$ lists during each node split or merge since each $top$ list can be computed using $O(b)=O(1)$ comparisons. As a result, node splits and merges cost $O(1)$ time and updating the score of points in a layer or altering its structure costs $O(\log n)$ time. This brings the total insertion cost down to $O((\frac{\log n}{\log\log n})^{2}+k\log n)$ amortized time.

This change also implies that at the time of a query, each $(a,b)$-tree's root only stores $1$ element with the highest score in the layer and as a result we can no longer directly apply Lemma \ref{lemma:TopKDomLemma}. To overcome this we build a Strict Fibonacci Heap~\cite{BrodalLT12} by inserting each point with the highest score from each layer. By querying the heap we are able to find (and delete) the top-$1$ dominating point. After deleting a point $p$ (belonging in a layer $L$) from the heap, we have to replace it by the point of $L$ with the next highest score. This point can be found by querying $L$'s tree for $p$. Due to the definition of $top$ lists, the point with the next highest score in $L$ is guaranteed to be amongst the $O(b)$ $top$ lists of each node in the search path. We insert all $O(b\log n)=O(\log n)$ such points in the heap and repeat the process until $k$ points have been deleted from the heap. In order to not have any duplicate points in the heap, we also employ a marking process. Deleting a point from the heap requires $O(\log n)$ time, while adding $O(\log n)$ points also requires $O(\log n)$ time. Since there aren't any duplicate points in the heap and the process is repeated $k$ times, the query phase of the algorithm requires $O(k\log n)$ time. The discussion of this section can be summarized in the following theorem:

\begin{theorem} \label{thm:sd}
Given a set $\mathcal{S}$ of $n$ $2$-dimensional points, we can support update operations in $O((\frac{\log n}{\log\log n})^{2}+k^{2}\log n)$ amortized time and top-$k$ dominating queries in $O(k)$ time. Alternatively, we can support update operations in $O((\frac{\log n}{\log\log n})^{2}+k\log n)$ amortized time and top-$k$ dominating queries in $O(k\log n)$ time.
\end{theorem}

\section{Dynamic Top-$k$ Dominating Points} \label{section:DynamicTopKDom}

The algorithms presented so far only support insertions due to the fact that all operations could be restricted in the first $k$ layers of maxima. However, assume the deletion of a point $p$ in layer $L_k$. It is possible that some points from $L_{k+1}$ might have to be inserted in $L_k$ as a result of them not being dominated by any other point in $L_k$ apart from $p$. This brings a cascading of restructuring operations since some of the points in $L_{k+2}$ might have to be inserted in $L_{k+1}$. Thus, a deletion operation may reach the last layer of $\mathcal{S}$ in the worst case.

Our algorithm for semi-dynamic settings can be extended to fully dynamic settings through the use of the global rebuilding technique~\cite{OvermarsL81}. More specifically in an update operation, instead of manipulating only the first $k$ layers we perform score updates and layer restructuring operations in the first $k+\sqrt{n}$ layers. A deletion of an existing point can be defined in a similar way to the insertion of a point with each layer requiring either score updating or restructuring. Since we stop restructuring operations on a predefined point, after the $i$-th deletion the $(k+\sqrt{n}-i+1)$-th layer will have become invalid. As a result, after $\sqrt{n}$ deletions, only the first $k$ layers remain valid and at that point we rebuild the entire layers of maxima data structure. We also recompute the score of each point and reconstruct the $(a,b)$-trees. Further details and the proof of the following theorem can be found in Appendix~\ref{app:global}.

\begin{theorem} \label{thm:fd}
Given a set $\mathcal{S}$ of $n$ $2$-dimensional points, we can support update operations in $O(\sqrt{n}(\frac{\log n}{\log\log n})^{2}+(k+\sqrt{n})k\log n)$ amortized time and top-$k$ dominating queries in $O(k)$ time. Alternatively, we can support update operations in $O(\sqrt{n}(\frac{\log n}{\log\log n})^{2}+(k+\sqrt{n})\log n)$ amortized time and top-$k$ dominating queries in $O(k\log n)$ time.
\end{theorem}

\section{Conclusions and Future Work} \label{section:ConclusionsFutureWork}
In this work we proposed algorithms for answering semi-dynamic and fully dynamic top-$k$ dominating queries where $k$ is a parameter that is fixed between queries. The algorithms we described offer asymptotic guarantees for both their time and space cost and they are the first to do so. An interesting future work direction would be to lower the update cost for the full dynamic algorithms by avoiding the global rebuilding technique. Another research direction would be to provide solutions for this problem appropriate for secondary memory.


%
%
\bibliographystyle{splncs}
\bibliography{references}

\newpage
\appendix

\section{Reporting Lemma}

We use the following lemma from \cite{Frederickson82}:
\begin{lemma}\label{lemma:TopKDomLemma}
    Let $A_1,\ldots,A_m$ be arrays of values from a totally ordered set such that each array is sorted.
    Given an integer $L \leq \sum_{i = 1}^m {\left| {{A_i}} \right|} $, there is a comparison-based algorithm that finds in $O(m)$ time a value $\tau$ that is greater than at least $L$ but at most $O(L)$ values in $A_1\cup\ldots\cup A_m$.
\end{lemma}

This lemma forms the basis in allowing us to efficiently find the $k$-th point with the highest score out of a collection of ordered lists and is used in Sections \ref{section:SemiDynTopKDom} and \ref{section:DynamicTopKDom}.

\section{Proof of Lemma~\ref{lemma:FirstKLayers}} \label{app:lemma1}

\textbf{Lemma} \textit{The top-$k$ dominating points of $\mathcal{S}$ are found in the first $k$ layers of maxima of $\mathcal{S}$.}

\begin{proof}
If $\mathcal{S}$ has only $k$ or less layers of maxima, the lemma obviously holds. Otherwise, assume that a point $p$ belongs in the $i$-th layer of maxima, where $i\geq k+1$. There are at least $i-1$ points dominating $p$ and due to Equation \ref{eq:DomScoreProperty} all of them have a larger score than $p$. As a result, $p$ is not included in the top-$k$ dominating points of $\mathcal{S}$.\qed
\end{proof}

\section{Proof of Lemma~\ref{lemma:AugABTreeTotalSpace}} \label{app:lemma2}

\textbf{Lemma} \textit{The total space required by an augmented $(a,b)$-tree storing $m$ points is $O(m)$.}

\begin{proof}
All the nodes with height lower than $\log_{b}k$ only store $O(1)$ additional information so their total space cost is $O(m)$. There are $O(m/k)$ nodes with height higher than or equal to $\log_{b}k$ each augmented with a $k$-sized list. The total space cost of this part of the data structure is $O(m/k)\times O(k)=O(m)$. As a result, the total space cost of the entire data structure is $O(m)$.\qed
\end{proof}

\section{Proof of Lemma~\ref{lemma:construction}} \label{app:construction}

\textbf{Lemma} \textit{The construction of an augmented $(a,b)$-tree over $m$ points can be carried out in $O(m\log{k})$ time, where $k$ is a user-defined parameter.}

\begin{proof}
In order to construct the leaf-oriented augmented $(a,b)$-tree we follow a bottom-up approach and assume that the input points are sorted according to their dimensions. The augmented $(a,b)$-tree is constructed in a similar way to a typical $(a,b)$-tree with an additional issue. At first, the nodes of the augmented $(a,b)$-tree are constructed by scanning the input points, creating the leaves and then recursively creating the inner nodes from bottom to top. Each node is only visited once so the procedure up to this point requires $O(m)$ time.

The last step is to compute the $top$ lists for all nodes with $height_v \geq \log_{b}{k}$. For each node $v$ with $height_v > \log_{b}{k}$, the $top(v)$ list must be computed from the $top$ lists of $v$'s children. By simultaneously traversing the $O(b)=O(1)$ $top$ lists of $v$'s children we can compute $top(v)$ in $O(k)$ time. There are $O(m/k)$ nodes with $height_v > \log_{b}{k}$ and since this process is repeated for every node, the time required is $O(m/k)\times O(k)=O(m)$.

Finally, we compute the $top$ lists for each node $v$ with $height_v = \log_{b}{k}$. Since $v$'s children are not augmented with $top$ lists we follow a different approach. We sort all the points found in $v$'s subtree\footnote{There are $k$ points in $v$'s subtree since $height_v = \log_{b}{k}$} in $O(k\log{k})$ time and store them in $top(v)$. There are $O(m/k)$ nodes with $height_v = \log_{b}{k}$ and thus this step requires $O(m\log{k})$ total time.

The lemma follows by the above discussion.
\end{proof}

\section{Global Reconstruction for the Fully Dynamic Case} \label{app:global}

At the beginning, we compute the dominance score of each point in $\mathcal{S}$ using the $2$-dimensional range counting data structure. This data structure does not need to be rebuilt since it is not related to the layers of maxima of $\mathcal{S}$.

The next step is to construct the layers of maxima for all the points in $\mathcal{S}$. This can be performed using the in-place algorithms proposed in \cite{BlunckVAlgorithmica10}. Afterwards, we build an augmented $(a,b)$-tree on the points of each layer of maxima.

Computing the dominance score of all the points in $\mathcal{S}$ using \cite{HeMunroWADS11} requires $O(n(\frac{\log n}{\log\log n}))$ time and $O(n)$ space. The construction of all the layers of maxima can be done in $O(n\log n)$ time and $O(n)$ space \cite{BlunckVAlgorithmica10}. Constructing the augmented $(a,b)$-tree for all layers-of-maxima requires $O(n\log k)$ time and $O(n)$ space.

We perform the global rebuilding step once in every $\sqrt{n}$ updates. A update up to the $(k+\sqrt{n})$-th layer, requires $O((\frac{\log n}{\log\log n})^{2}+(k+\sqrt{n})k\log n)$ amortized time. We perform $\sqrt{n}$ such updates and then we globally rebuild the data structures in $O(n(\frac{\log n}{\log\log n})^{2})$ time so the amortized time for an update is $O(\sqrt{n}(\frac{\log n}{\log\log n})^{2}+(k+\sqrt{n})k\log n)$. Finally, the methods described in Section \ref{subsection:ReduceUpdateCost} can also be applied in the full dynamic setting.

\end{document}